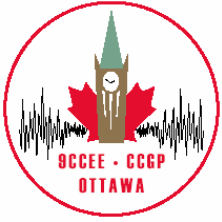

# SEISMIC RISK SCENARIO IN GRENOBLE (FRANCE) USING EXPERIMENTAL DYNAMIC PROPERTIES OF BUILDINGS

C. Michel[1] and P. Guéguen[2]

### ABSTRACT

Assessing the vulnerability of a large set of buildings using sophisticated methods can be very time consuming and at a prohibitive cost, particularly for a moderate seismic hazard country like France. We propose here a low-cost analysis using an experimental approach to extract the elastic behaviour of existing buildings. An elastic modal model is proposed for the different types of building tested in Grenoble (France) thanks to their experimental modal parameters (resonance frequencies, modal shapes and damping), which are estimated using ambient vibrations surveys. Sixty buildings of various types were recorded. The building integrity is then calculated considering an accelerogram scenario provided by seismologists as input and considering an integrity threshold based on the FEMA inter-storey drift limits. Even if the level of damage remains unknown, we conclude that masonry buildings undergo more damage (70% of buildings damaged) than RC buildings. Finally, extracting modal parameters from ambient vibration recordings allows us to define, for each class of building, its ability to support seismic deformation in case of earthquake.

### Introduction

Performing a seismic risk scenario at the urban scale requires both a realistic input earthquake and an overview of the building stock vulnerability. Even if this subject is still under research, the seismologists are now able to provide realistic ground motion (Causse et al. 2006) using empirical or numerical approaches with their variability. Engineers often do not pay attention to their improvements in the knowledge of strong ground motions. The vulnerability at urban scale is often represented by damage probability matrices based on observed damage (GNDT 1986), which link the intensity of the ground motion to the damage grade for different types of buildings. At the scale of the building, engineers model the structure and compare a demand (hazard) to a capacity (vulnerability) to determine the performance (damage grade) of the building (RISK-UE 2003). We propose here a hybrid method using simple models of buildings based on experimental parameters extracted from ambient vibrations and standing for different types of buildings found in Grenoble city (France), one of the most exposed French cities to seismic hazard. We recorded ambient vibrations in approximately 60 various buildings and extracted their modal parameters that were used in a simple elastic modal model. Then, using an accelerogram scenario provided by seismologists, we compute the elastic drift along the building and compare it to the classical elastic threshold for the materials of the 60 buildings (RC and masonry) to conclude on the integrity of building after shaking.

---

[1]PhD student, Laboratoire de Géophysique Interne et Tectonophysique (LGIT), University of Grenoble, FRANCE
[2]Researcher, LGIT – LCPC, University of Grenoble, FRANCE

## The Grenoble City and its seismic hazard

### Building inventory

Grenoble is a 400,000 inhabitants city located in the centre of the French Alps. It was only a small town during the centuries from the Roman Empire to the 19th century, enclosed in its walls. It grew rapidly after World War II, especially because of a great increase of immigration. The typology of the building stock (Building Database Typology - BDT Grenoble) was designed in the frame of the Vulneralp project (Guéguen and Vassail 2004). Six different types of concrete and thirteen types of masonry buildings are proposed. In this particular study we group types for the sake of simplicity. We distinguish 2 types of concrete buildings: before 1950 (B1 to B3 on the BDT) and after 1950 (B4 to B6). Although the differences in the masonry types will be taken into account in the future, we do not distinguish them in this paper.

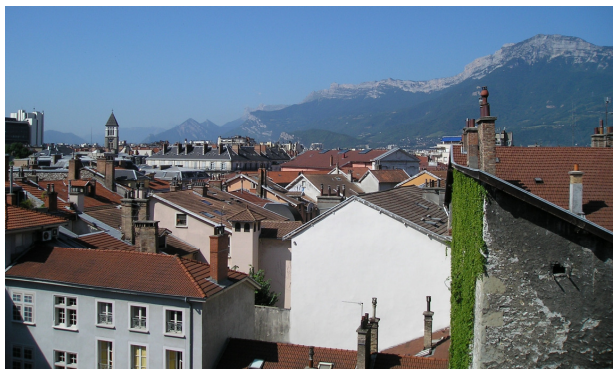

Figure 1. The Grenoble downtown is constituted by a large majority of old masonry buildings and few RC structures after 1950's.

### Seismic hazard

The seismicity of this Grenoble area is moderate (Fig. 2). The strongest earthquakes of the last centuries are the 1962 Ml=6.2 Corrençon earthquake and in the last decade, the 1996 Ml=4.9 Annecy earthquake, with Ml the local magnitude. These earthquakes caused slight damages in near field but Grenoble has never suffered damage. A noticeable instrumental seismicity exists in the Belledonne massif leading to discover the hidden Belledonne border fault (BBF) (Thouvenot et al. 2003), only a few kilometers away from Grenoble. According to experts (ESG 2006), the maximum expected event on this fault is Ml=5.5 located 15 km away from the city. In addition, the city lies on a deep sedimentary basin producing strong site effects (ESG 2006).

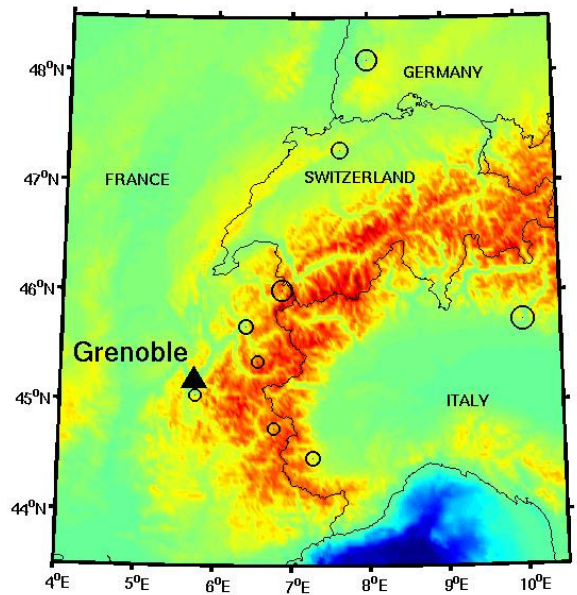

Figure 2. Location of Grenoble and epicenters of the 9 strongest earthquakes recorded in Grenoble between November 2004 and December 2005 (Magnitude 5.5 to 3.1)

**Experimental Survey of the Building Stock**

**Ambient vibration recordings**

In 2005 and 2006, two ambient vibration surveys were carried out in Grenoble. They consisted of recording ambient vibrations at each story for approximately 60 buildings of various types. We used a Cityshark II ambient vibrations station (Chatelain et al. 2006) that allows the simultaneous recording of 6 3D Lennartz 5s sensors. These sensors have a flat response in velocity between 0.2 and 50 Hz. The length of the recording was 15 min at a 200 Hz sampling frequency so that it allows a precise measurement between approximately 0.5 and 25 Hz which is the classical frequency range of engineering structures. For buildings more than 6 floors, several datasets were recorded keeping a reference sensor at the top. Moreover, all the structural characteristics of the buildings were collected including dimensions, age, construction material, type following the BDT description.

**Processing of data**

The modal parameters of each building (resonance frequencies, modal shapes and damping) are calculated using the Frequency Domain Decomposition method (FDD) (Brincker et al. 2001). The principle of this method is first to calculate the Power Spectral Density (PSD) matrices, i.e. the Fourier Transforms of the correlation matrices between each simultaneous recording. Then it performs the Singular Value Decomposition of these matrices at each frequency. As only 1 (or 2) mode has energy at one particular frequency, the first (or 2 first) singular value shows peaks corresponding to the structural modes and the corresponding singular vectors stands for the modal shape. The damping is the logarithmic decrement of the mode's bell represented in the time domain. In this study, we did not put the stress on damping which traduces a complex process. For sake of simplicity, we choose a constant damping ratio of 4%. After this processing, 51 buildings remained, because the results were not clear for the other buildings.

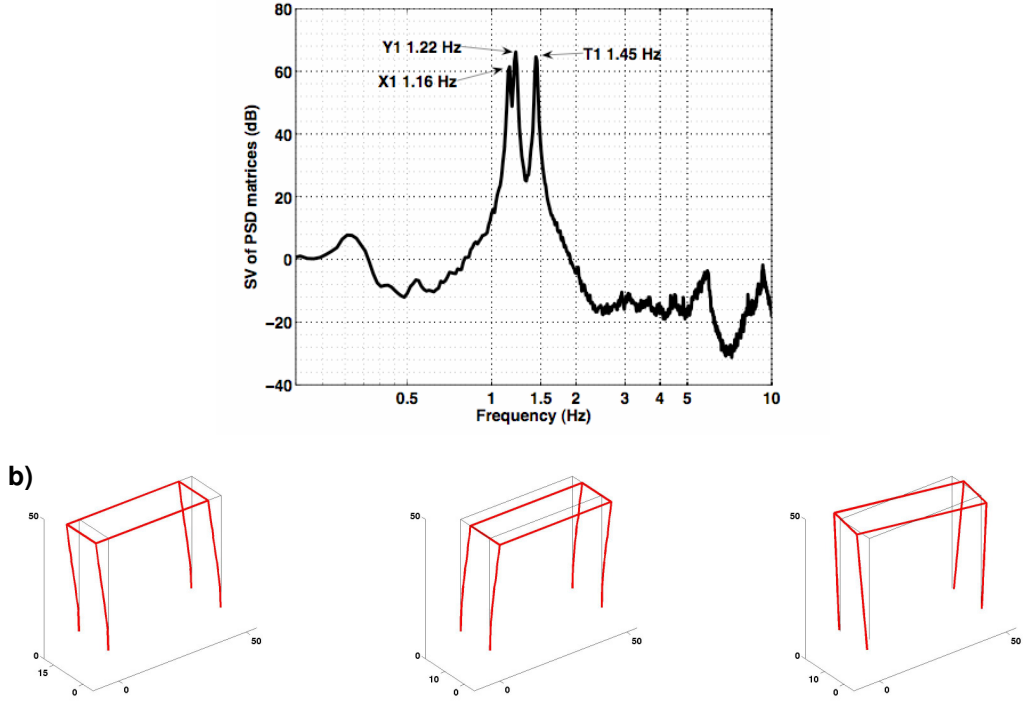

Figure 3. a) Example of spectrum (first singular value of the PSD matrices) for one of the 60 buildings (Grenoble City Hall) under ambient vibrations; b) Corresponding modal shapes, from left to right: longitudinal bending (1.16 Hz), transverse bending (1.22 Hz) and torsion (1.45 Hz)

**Modal Model**

**Drift computed for an earthquake scenario**

The modal parameters obtained under ambient vibrations are unscaled, i.e. it is not possible to deduce the amplitude of the building motion knowing the input excitation. Therefore, we need a physical model that would integrate these parameters and some hypothesis. Because we can consider the mass of each story as mostly concentrated at its floor, we assumed a lumped-mass modelling for this structure. In this case, the Duhamel integral (Clough and Penzien 1993) gives us the elastic motion of the structure at each floor $\{U(t)\}$ knowing only the mass of the stories $[M]$, the vibration modes ($[\Phi]$ the modal shapes, $\{\omega\}$ the frequencies and $\{\xi\}$ the damping ratios) and the ground motion $U_s''(t)$:

$$\{U(t)\} = [\Phi]\{y(t)\} \qquad (1)$$

with $\forall j \in [1, N]$ $\quad y_j(t) = \dfrac{-p_j}{\omega'} \displaystyle\int_0^t U_s''(\tau) e^{-\xi_j \omega_j (t-\tau)} \sin(\omega'(t-\tau)) d\tau$, (2)

$\omega_j'^2 = \omega_j^2(1-\xi_j^2)$ and $p_j = \dfrac{\{\Phi_j\}^T[M]\{1\}}{\{\Phi_j\}^T[M]\{\Phi_j\}}$ the participation factor of mode j. (3)

We assume here a mass of 1000 kg/m² for each floor (standard values for French buildings) and we consider that only the two first bending modes in each direction provided energy, neglecting the torsion modes for sake of simplicity. It is then possible to compute the motion at each floor for any deterministic earthquake scenario. This is of course a linear model, which suits only for moderate motions.

Nevertheless, as mentioned in Boutin et al. (2005), elastic modelling can be used to detect whether the building reaches the post-elastic state or not.

**Maximum drift**

The Federal Emergency Management Agency published (FEMA 386 2000) maximum inter-storey drifts for 3 different performances of buildings: Immediate Occupancy, Life Safety and Collapse Prevention. We consider that our elastic model is not valid anymore after the Immediate Occupancy threshold. We consider this threshold for different types of building in the FEMA document (Tab. 1). If at least one story has a maximum drift greater than this threshold, the building is considered damaged. Of course this damage can be slight or heavy but our elastic model is not able to determine it. In addition, RC-buildings are more able to sustain plastic deformation on the contrary to masonry buildings that collapse rapidly after the elastic limit is reached. It means that this damage parameter (elastic limit) may not be relevant for strong earthquakes for which buildings may collapse. However for moderate earthquakes, which are the purpose in this paper, it may predict the zones that should be retrofitted after (or before) the earthquake.

Table 1. Maximum inter-storey drift for immediate occupancy (IO) according to FEMA 386 for different types of buildings.

| Type | | BDT Type | Max Drift IO |
|---|---|---|---|
| **Reinforced Concrete** | Walls with shear dominating | B4,B5,B6 | $4.10^{-3}$ |
| | Beams and columns | B1,B2, B3 | $3.10^{-3}$ |
| **Masonry** | Unreinforced | M1-M13 | $10^{-3}$ |

## Earthquake scenario in Grenoble

**Input accelerogram**

The earthquake scenario corresponds to a magnitude Ml=5.5 earthquake with its epicenter 14.5 km away from Grenoble. Using the empirical Green's function method, Causse et al. (2006) computed the strong ground motion produced by this earthquake in the center of the city. Their method includes source variability, propagation and site effects. Due to site and source effects, the ground motion computed for this earthquake is quite strong. The median value reaches 0.4 g so that a large part of the buildings of the city, which are not designed following the national seismic code, can be damaged. For this reason, we choose a median minus a standard deviation motion. This motion is still higher than the design code (1.5 m/s$^2$) reaching 2.68 m/s$^2$.

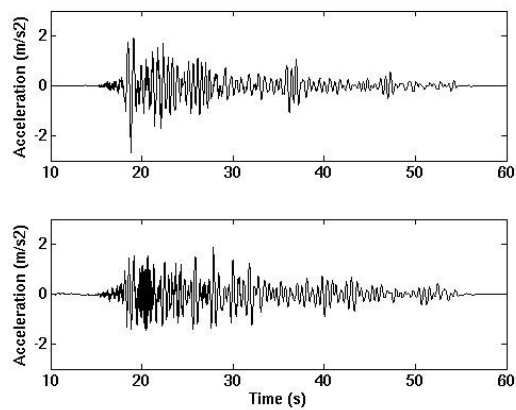

Figure 4. Input accelerogram (Causse et al. 2006) used for the scenario (top: NS direction ; bottom: EW direction). It simulates a Ml=5.5 at 14.5 km of Grenoble including source, propagation and site effects using the Empirical Green's function method.

**Damage forecast**

For this earthquake, 27% of the 15 RC shear wall buildings, 43% of the 7 RC shear beam buildings and 72% of the 29 unreinforced masonry buildings reached the first damage point. In the case of the RC buildings, this damage point is often reached at only one story. Fig. 5 shows the example of 2 RC shear beam buildings. The tallest reaches this damage point only at the top floor in the 2 directions whereas the other one, despite a clear soft story, is not damaged. The case of the masonry buildings is very different. Most of the tall masonry buildings reach the threshold at each story. Only the low-rise buildings are not damaged.

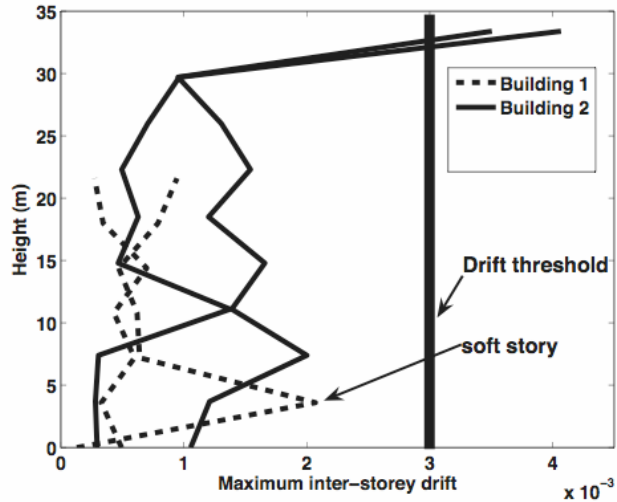

Figure 5.   Maximum drift along the stories for two RC shear beam buildings in each direction. Only the tallest one is damaged at the top floor.

## Conclusions

We showed here the basis of a method to perform risk scenario at the urban scale. This work is only preliminary and many points can and will be improved. The main point is that we used experimental data to build simple models of buildings. The data recordings and computations are quite fast and low cost and the computed modal parameters take into account the unknown distribution of stiffness in each building. The disadvantage is that it is then very difficult to have a statistical idea of the variability of the buildings among one type.

The damage parameter we chose (elastic limit) is only relevant for moderate earthquakes. The ductile behavior of RC and masonry buildings is very different and is not taken into account here.
We showed that, especially for masonry buildings, the height was a key-parameter. Indeed, the input accelerogram have a prevalent frequency around 2Hz, a frequency of many 8 to 10-story buildings of the city, because of thin layers resonance in the basin.

## Acknowledgments

This work is part of the Rhône-Alpes Region project VulneRAlp (Seismic vulnerability at the scale of a city in Rhône-Alpes, application to Grenoble).